\documentstyle[aps,epsfig,floats]{revtex}
\draft
\input epsf

\newcounter{fig}
\newcommand{\gsim}{\lower.7ex\hbox{$\;\stackrel{\textstyle>}{\sim}\;$}}
\newcommand{\lsim}{\lower.7ex\hbox{$\;\stackrel{\textstyle<}{\sim}\;$}}

\begin{document}
%
\twocolumn[\hsize\textwidth\columnwidth\hsize\csname
@twocolumnfalse\endcsname
%

\title{Defect Formation Rates in
  Cosmological First-Order Phase Transitions}
 
\author{Matthew Lilley\footnotemark and Antonio Ferrera\footnotemark}
 
\address{\footnotemark[1] Department of Applied
  Mathematics and Theoretical Physics, University of Cambridge,\\
  Cambridge CB3~9EW, United Kingdom } \address{\footnotemark[1]
  Institut f\"ur Theoretische Physik der Universit\"at Heidelberg\\ 
  Philosophenweg 16, D-69120 Heidelberg, Germany}
 
\maketitle
 
\pagenumbering{arabic}
 
\begin{abstract}In cosmological first-order phase transitions, the
  progress of true-vacuum bubbles is expected to be significantly
  retarded by the interaction between the bubble wall and the hot
  plasma.  It has been claimed that this leads to a significant
  reduction in the number of topological defects formed per bubble, as
  a result of phase equilibration between bubbles.  This claim has
  been verified for spontaneously-broken global symmetries.  We
  perform a series of simulations of complete phase transitions in the
  $2+1$-dimensional $U(1)$-Abelian Higgs model, for a range of bubble
  wall velocities, in order to obtain a quantitative measure of the
  effect of bubble wall speed on the number density of topological
  defects.  We find that the number of defects formed is i)
  significantly lower in the local than the global case and ii)
  decreases exponentially as a function of wall velocity.  Slow-moving
  bubbles also lead, however, to the nucleation of more bubbles before
  the phase transition is complete.  Our simulations show that this is
  in fact the dominant effect, and so we predict {\em more} defects
  per unit volume as a result of the sub-luminal bubble wall terminal
  velocity.

\end{abstract}
]
 
\footnotetext{~\vspace{-.8cm}~}
\footnotetext{\footnotemark[1]Electronic address: {\tt
    lilley@thphys.uni-heidelberg.de}}
\footnotetext{\footnotemark[2]Electronic address: {\tt
    ceefp44@pinar2.csic.es }} 
 
\renewcommand{\thefootnote}{\arabic{footnote}}
 

\section{Introduction} \label{sect:intro}

As the early Universe expanded and cooled, it is expected to have
undergone a series of phase transitions at each of which a symmetry
which had been thermally-restored became spontaneously-broken.  At
each of these phase transitions, depending on the nature of the
symmetry-breaking involved, topological defects may or may not have
formed \cite{kibble}.  Phase transitions are labelled first- or
second-order according to whether the field-space location of the true
vacuum state changes discontinuously or continuously as the critical
temperature is crossed.  First-order phase transitions, with which
this work is concerned, proceed by bubble nucleation and expansion.
Depending on the details of the theory, when at least $(4-n)$ bubbles
collide, an $n-$dimensional topological defect may form in the region
between them.  In this paper, we simulate a series of complete
cosmological first-order phase transitions in order to quantify the
initial density of topological defects.

We take as our model the simplest spontaneously-broken gauge symmetry:
the Abelian Higgs model, which has a local $U(1)$ symmetry, with
Lagrangian
\begin{equation} \label{ssbscalar}
{\mathcal L} =  {( {D}_{\mu} \Phi )}^{\dagger}{( {D}_{\mu} \Phi )} - {1 \over
  4}F_{\mu \nu}F^{\mu \nu}- V(\Phi^{\dagger}\Phi),
\end{equation} 
where ${D}_{\mu} \Phi = { \partial }_{\mu}\Phi - ie{A}_{\mu}\Phi$ and
$F_{\mu \nu} = { \partial }_{\mu}{A}_{\nu} - { \partial
  }_{\nu}{A}_{\mu}$.  The detailed form of the effective potential
$V(|\Phi|)$ will depend upon the particular particle-physics model
being considered, but in order to be able to study the generic
features of a first-order phase transition we shall take $V$,
following \cite{ferrera_melfo} and \cite{lilley_bubble}, to be
\begin{equation} \label{potentialeqn}
V(\Phi) = \lambda \Bigl[{\frac{|\Phi |^{2}}{2}}(|\Phi | -
  \eta)^{2} - \frac{\varepsilon}{3}{\eta}|\Phi |^{3}\Bigr],
\end{equation}
where in a realistic model, $\varepsilon = \varepsilon (T) \propto
(T_c - T)$.  The potential $V$ has a local minimum false-vacuum state
at $\Phi = 0$ which is invariant under the $U(1)$ symmetry, and global
minima true-vacuum states on the circle $|\Phi |= \rho_{tv}\equiv
\left(\eta / 4\right)(3 + \varepsilon + \sqrt{1 + 6\varepsilon +
  \varepsilon^2})$ which possess no symmetry.  The dimensionless
parameter $\varepsilon$ is responsible for lifting the degeneracy
between the two sets of minima -- the greater $\varepsilon$, the
greater the potential difference between the false- and true-vacuum
states, and hence the faster the bubbles will accelerate.  This model
admits string-solutions, or in the $2+1$-dimensional case considered
here, vortices.

By making the field and coordinate transformations
\begin{eqnarray} \label{transformations}
\Phi \longrightarrow \phi &=& \eta  \Phi \\
{\bf x} \longrightarrow {\bf x'} &=& \frac{{\bf x}}{\sqrt{\lambda}\eta} \\
t \longrightarrow t' &=& \frac{t}{\sqrt{\lambda}\eta}
\end{eqnarray}
it is possible to set $\lambda$ and $\eta$ to unity, so that the
potential is parametrised only by $\varepsilon$, and hereafter we
shall use these transformed variables.

The bubble nucleation rate per unit time per unit volume and the
field-space profile of the bubble wall are obtained from the `bounce'
(i.e. least action) solution of the Euclidean action \cite{coleman}.
In $d+1$-dimensions the modulus $\rho$ of the field $\Phi$ is given by
the solution of the equation
\begin{equation} \label{bounce}
{d^2 \rho \over dr^2} + \left({d\over r}\right) {d \rho \over dr} -
V'(\rho) = 0,
\end{equation}
with boundary conditions
\begin{eqnarray}
\lim_{r \to \infty}\rho(r) &=& 0  \\
\frac{d \rho(r)}{d r}\Bigg|_{r=0} &=& 0.  
\end{eqnarray}

Ignoring quantum fluctuations, the phase $\theta$ is constant within
each bubble, and uncorrelated between spatially-separated bubbles.
Any non-zero gauge fields in the nucleation configuration will make a
contribution to the action and hence the nucleation of bubbles with
non-zero gauge fields is exponentially suppressed.  When three or more
bubbles collide, a phase-winding of $2{\pi}n$ can occur around a
point, which by continuity must then be at $\Phi = 0$.  In three
spatial dimensions, this topologically-stable region of high-energy
false vacuum is string-like -- a cosmic string.  If the phases of the
bubbles were uniformly-distributed, it is simple to calculate that at
every three-bubble collision, there is a probability $p=0.25$ that the
arrangement of phases would be such as to generate the winding
required to form a defect.  We will use this value of $p=0.25$ as our
benchmark throughout this work. This description of defect formation
and the estimation of the initial defect density is known as the
Kibble mechanism, and the purpose of this work is to investigate to
what extent its predictions are modified by the early-Universe
environment in which cosmological phase transitions take place.

Although two-dimensional (or domain walls) \cite{zeld_domain} and
pointlike defects (monopoles) \cite{zeldovich,preskill} would dominate
the energy density of the Universe, one-dimensional `cosmic strings'
are expected to reach a scaling solution (where the ratio of the
string energy density to that of the background remains constant), and
thus need not be pathological.  For many years, cosmic string-seeded
perturbations were considered a viable candidate for the primordial
density fluctuations which led to the formation of structure in the
Universe.  The failure of cosmic string-induced CMB anisotropy spectra
\cite{turok} to match observations \cite{boom,maxima} however, has
laid rest to this theory.  This does not mean though that cosmic
strings did not form, or that if they did, they are
cosmologically-irrelevant.  

Cosmic strings may be either closed loops, or infinitely-long strings
(see \cite{shellard_vilenkin} or \cite{hindmarsh_kibble} for cosmic
string reviews).  Closed loops will decay via gravitational radiation,
forming a gravitational-wave background which could be detectable in
the planned experiments LIGO and LISA. Cosmic strings have been cited
as being responsible for baryogenesis \cite{brand,perkins}, if they
trap a Grand Unified gauge field with baryon-number violating
interactions in their core.  They could cause the highest energy
cosmic rays \cite{bhatt}, whose existence has largely defied
explanation -- standard methods of ray production such as quasars do
not appear to be able to generate such events.  When string loops
double back on themselves, forming a `cusp', a jet of ultra-high
energy accelerated particles is emitted, which could explain the
origin of Gamma Ray Bursts \cite{grb}.  It has even been proposed that
cosmic strings may constitute the dark matter \cite{bucher} -- a
network of non-Abelian (in order to prevent a scaling solution being
reached) strings would have negative pressure and could therefore
explain the why the Universe appears to be accelerating.

In order to be able to assess the significance of cosmic strings in
the evolution of the early Universe, it is important to be able to
estimate the initial defect density accurately.  If the phase
transition at which they formed was first-order, this would depend on
how the phases between two or more bubbles interpolate after
collision.  In particular, although strings are in general formed when
three or more bubbles collide, a simultaneous three-bubble collision
is unlikely -- one would expect in general two-bubble collisions, with
a third, or fourth bubble colliding some finite time later.  If the
phase inside a two-bubble collision is able to equilibrate quickly,
and before a third bubble arrives, there may be a strong suppression
of the initial string density.  The effect of phase equilibration on
the initial defect density was first investigated by Melfo and
Perivolaropoulos \cite{MP}.  They found a decrease of less than $10\%$
(i.e. $0.22 < p < 0.25$), in models which possess a global symmetry
and with bubbles which, upon nucleation, accelerated up to the speed
of light.

The above description of defect formation, however, ignores any effect
that the hot-plasma background may have on the evolution of the Higgs
field, which may be significant in the early Universe.  Real-time
simulations \cite{laine} and analytic calculations \cite{bubble} for
the (Standard Model) electroweak phase transition predicted that the
bubble wall would reach a terminal velocity $v_{\rm ter} \sim 0.1c$.
Recent calculations for the Minimal Supersymmetric Standard Model,
where there are many more particles, and hence more interactions and
hence a more viscous plasma, show that the bubble wall velocity could
be as low as $v_{\rm ter} \sim 10^{-3} c$ \cite{john}.  The reason for
this is simple: outside the bubble, where the ($SU(2)\times U(1)$)
symmetry remains unbroken, all fields coupled to the Higgs are
massless, acquiring their mass from the vacuum expectation value of
the Higgs in the spontaneously-broken symmetry phase inside the
bubble.  Particles outside the bubble without enough energy to become
massive inside bounce off the bubble wall, retarding its progress
through the plasma.  The faster the bubble is moving, the greater the
momentum transfer in each collision, and hence the stronger the
retarding force.  Thus a force proportional to the bubble-wall
velocity appears in the effective equations of motion.
 
Ferrera and Melfo \cite{ferrera_melfo} studied bubble collisions in
such an environment, for theories which possessed a global symmetry,
and found that decaying phase oscillations occur inside a two-bubble
collision, which was claimed would lead to a suppression of the defect
formation rate.  Kibble and Vilenkin \cite{kibble_vilenkin} studied
phase dynamics in collisions of undamped bubbles in models with a
local symmetry, and found, analytically, a different kind of decaying
phase oscillation.  When the finite conductivity of the plasma was
included, these oscillations were found not to occur.  Davis and
Lilley \cite{lilley_bubble} extended this work, to include slow-moving
bubbles in local-symmetry models. Simulating the collision and merging
of two bubbles, it was found that the phase oscillations were
suppressed -- the (gauge-invariant) phase difference between two
bubbles equilibrated very rapidly upon collision (and even more
rapidly when the conductivity of the plasma was taken into account),
with the implication that there would be a suppression of defect
formation in slow-moving bubbles in gauge theories as well.  This
claim was illustrated by an example of an undamped 3-bubble collision
which led to the formation of a defect, but when identical initial
conditions were evolved in a damped environment, no defect formed.
 
Ferrera \cite{ferrera98} confirmed the hypothesis of
\cite{ferrera_melfo} by performing a series of simulations of complete
global-symmetry phase transitions at different values of the bubble
wall terminal velocity.  However, quantitative analysis of the defect
formation rate in the most realistic scenario cosmologically -- a
gauge-theory phase transition where the bubbles are slowed
significantly by the plasma (as might be expected at the electroweak-
or GUT-scales) -- has not been studied.

In this paper we present the results of our investigations into the
effect of the sub-luminal bubble wall velocity in gauge theory phase
transitions.  We have simulated a series of complete phase
transitions, from the nucleation of the first bubble in an entirely
symmetric background until $>95 \%$ of the available space having been
converted into the broken-symmetry phase.  We have done this (as in
\cite{ferrera98}, but in a local rather than global symmetry) for a
range of values of the bubble wall velocity in order to obtain a
qualitative measure of the dependence of $n_d$ on $v_{\rm ter}$.  Our
results (which are displayed in Figure \ref{fig:sim_graph}) show that
the two hypotheses of \cite{lilley_bubble} {\em do} hold: for any
value of the bubble wall speed there are fewer defects in gauge theory
phase transitions than in global theory ones, and the number of
defects formed per bubble decreases with decreasing wall velocity.
For values of the wall velocity which are high compared with those
calculated for cosmological phase transitions, we find already almost
an order of magnitude fewer defects.  We also confirm the gauge-theory
analogue of the result of \cite{MP} -- if the bubbles move at the
speed of light, there is no significant reduction in the defect
formation rate due to phase equilibration.

As well as the possibility of phase equilibration and the suppression
of defect formation however, slow-moving bubbles will also require a
larger number of bubbles to be nucleated in order to complete the
phase transition. These two effects act in opposition, and our
simulations show, for the range of wall velocities considered, that
the former is dominant, i.e. the net effect of slow-moving bubbles is
that {\em more} defects per unit volume will be formed than one would
expect were the bubbles to move at the speed of light.

We should note in passing that we have ignored the effect of the
expansion of the Universe in our work.  This is a good approximation
for phase transitions which take place at late times, like the
electroweak phase transition.  At phase transitions which occur
earlier, however, the Hubble expansion may have a significant effect
on bubble and phase dynamics.  This topic deserves consideration on
its own.

\section{Simulating a Phase Transition}

In the previous section we have described how the sub-luminal terminal
velocity of the bubble walls causes the gauge-invariant phase
difference between two bubbles to equilibrate more quickly than in the
undamped case.  It has been demonstrated, qualitatively, how this can
lead to the suppression of topological defect formation
\cite{lilley_bubble}, and we have explained how this could be
cosmologically significant.  In order to set this work on firmer
footing, and be in a position to assess the cosmological consequences,
we would like now to be able to quantify these effects.

In particular, we would like to be able to investigate two of the
claims made in \cite{lilley_bubble}:

\begin{enumerate}
\item For a given wall velocity, fewer topological defects are formed
in local-symmetry models than in those with a global symmetry.

\item In models where a local symmetry is spontaneously broken, the
average number of defects formed per bubble nucleated is a decreasing
function of the bubble wall velocity.
\end{enumerate}

In order then to quantify the effect on defect formation probabilities
of the slow-moving bubble walls, we must perform many simulations of
`complete' phase transitions each involving the nucleation of many
bubbles.  Such an investigation was carried out for values of the
terminal velocity in the range $v_{\rm ter} = c$ to $v_{\rm ter}
\approx 0.05 c$.  This paper describes the procedure followed and
presents the results obtained.
 
In the standard \cite{vach_vilen} computer simulations of defect
formation, causally-disconnected points on the spatial lattice are
assigned randomly-generated relative phases (corresponding to the
centres of true-vacuum bubbles in a first-order phase transition, or
of domains in a second-order transition). The phase between sites is
then taken to vary on the shortest path on the vacuum manifold -- the
so-called geodesic rule.  Topological defects will then form wherever
this geodesic interpolation between sites generates a
topologically-nontrivial path in the vacuum manifold. For a
first-order transition this formalism corresponds to true vacuum
bubbles nucleating simultaneously, equidistant from all their nearest
neighbors. Consequently all collisions between neighboring bubbles
occur simultaneously, and the associated phase differences are simply
given by the differences in the initial assigned phases.

We, on the other hand, wish to investigate precisely the effect on the
defect formation probability of phase equilibration -- something which
is only possible as a result of the non-simultaneity of three-bubble
collisions.

The procedure described above was formulated as a consequence of the
limitations of computing power available at the time.  Now, however,
with the increased processor capabilities, we have the opportunity to
perform much more computationally-intensive simulations.  As a result,
instead of following the standard procedure, we have used a spatial
lattice several times smaller than the width of the bubble wall, and
evolve the discretised field equations in their entirety.  This is
necessary as it is the dynamics of the phase and gauge fields which
lead to the suppression of defect formation seen in
\cite{lilley_bubble}.

\section{Phase Transition Algorithm} \label{sect:algorithm}

We wish to simulate a first-order phase transition in the
Abelian-Higgs Model, which possesses a $U(1)$ gauge symmetry and has
equations of motion

\begin{eqnarray} \label{ab_higgs_rho} 
\ddot{\rho} - {\rho}'' - ({\partial}_{\mu}\theta -
e{A}_{\mu})^{2}\rho & = &-\frac{\partial V}{\partial \rho}\\
\label{ab_higgs_theta} \partial^{\mu}\bigl[\rho^{2}(\partial_{\mu}\theta
- eA_{\mu})\bigr] &=& 0\\
\label{ab_higgs_a}\ddot{A_\nu} - {A_\nu}'' -
{\partial}_{\nu}\left(\partial
 \cdot A \right) & = & -2e \rho^2 {\partial}_{\nu}\theta,
\end{eqnarray}   
where we will take the potential $V$ to be given by
\begin{equation} \label{ab_higgs_pot}
V(\Phi) = \lambda \Bigl[{\frac{|\Phi |^{2}}{2}}(|\Phi | -
  \eta)^{2} - \frac{\varepsilon}{3}{\eta}|\Phi |^{3}\Bigr],
\end{equation}   
with $\lambda$ and $\eta$ re-scaled to unity in the manner described
in the Introduction. As we have stated however, we would like to
investigate the behaviour of the phase in collisions of slow-moving
bubbles.  For a given theory, by considering the Boltzmann equations
for scattering off the Higgs field, it is possible to calculate the
terminal velocity of the bubble wall \cite{lmt}.  Since we are not
concerned here with the parameters of a specific particle-physics
model, we choose instead to use a single damping parameter $\Gamma$ to
model the interaction of the Higgs with the plasma.  In the
introduction we claimed that the plasma would introduce a term
proportional to the bubble-wall velocity into the equations of motion.
Since the phase $\theta$ of the Higgs field is not affected by the
effects described, we assume that the plasma couples only to the
modulus $\rho$.  We then have an effective equation of motion for
$\rho$
\begin{equation}
\label{rho_damp} \ddot{\rho} - {\rho}''  + \Gamma \dot{\rho} -
({\partial}_{\mu}\theta - e{A}_{\mu})^{2}\rho = -\frac{\partial
  V}{\partial \rho},
\end{equation}
replacing equation (\ref{ab_higgs_rho}).  Equations
(\ref{ab_higgs_theta}) and (\ref{ab_higgs_a}) remain unchanged.

A damping term of this form has been used by several authors
\cite{ferrera_melfo}, \cite{heckler}, \cite{lmt}, and has also been
derived from the stress-energy of the Higgs, assuming a coupling to
the plasma \cite{ignatius}.  Heckler \cite{heckler} estimates $\Gamma
\sim g_{W}^2 T_c$ for the electroweak phase transition, by comparing
the energy generated by the frictional damping with the pressure on
the wall due to the damping.
 
The effect of this damping term is that instead of accelerating up to
the speed of light, the bubble walls reach a terminal velocity $v_{\rm
  ter} < c$.  An exact relation between $\Gamma$ and $v_{\rm ter}$ can
be calculated \cite{lilley_bubble}, but the approximation $v_{\rm ter}
\sim 1 / \Gamma$ is sufficiently accurate for our purposes.

\begin{figure}
\begin{center}
  \epsfig{file=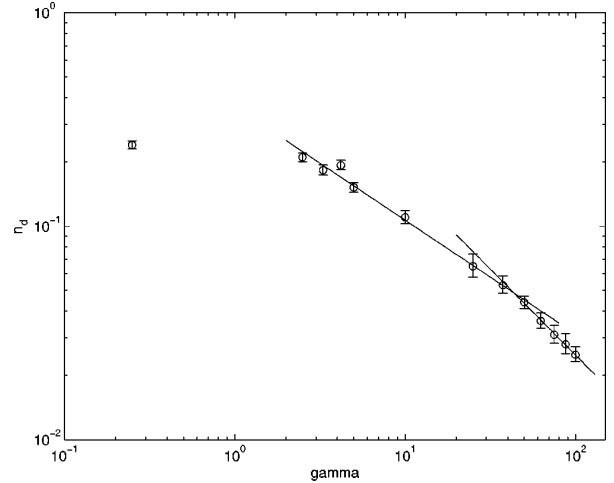 ,width=9cm}
\caption{Log-log plot of the number of defects formed per nucleated
bubble, $n_d$, vs the friction coefficient gamma (from
\protect\cite{ferrera98}).  The two solid lines correspond to the least
squares best fits for the two different regimes $n_d \propto
\gamma^{-\alpha_{1,2}}$.  The ``low'' friction regime has a slope of
$\alpha_1 = 0.53 \pm 0.05$, whereas in the high friction regime $n_d$
decays with $\alpha_2 = 0.8 \pm 0.3$.}\label{fig:ferrera}
\end{center}
\end{figure}

The results of Ferrera \cite{ferrera98} are shown in Figure
\ref{fig:ferrera} -- the number of defects formed per bubble
nucleated, $n_d$, decreases exponentially with decreasing bubble wall
velocity.  For terminal velocities $v_{\rm ter} \approx 0.01 c$, a
decrease of over an order of magnitude in the defect formation rate
(i.e. $p < 0.03$) was measured.

We will then follow a modified version of Ferrera's program (as in his
work using only 2+1 dimensions, again because of computational
constraints -- the gauge theory has many more degrees of freedom than
the global theory and thus requires significantly more memory and
computer time) of nucleating bubbles at random points in space and
time during the course of the simulation.  Our algorithm is as
follows:

\begin{enumerate}
\item for the specific parameters of the potential under
investigation, compute the bubble wall profile (numerically) from
equation (\ref{bounce}).

\item generate a population of time ordered bubble nucleation events
distributed randomly within some finite volume of 2+1 dimensional
space-time, assigning a random phase to each bubble.    

\item  start with an initial state for which the field is in the false
vacuum (i.e $\Phi = 0$) in all the simulation volume, and, assuming
that there is no primordial magnetic field, the gauge fields are
everywhere zero.

\item  at the beginning of each time step check whether there are any
bubbles to be nucleated at that time:  

\begin{enumerate}
\item if there are any bubble nucleation events, nucleate only those
  bubbles that would fall entirely in a false vacuum region and
  discard the rest (i.e., avoid superimposing new bubbles on regions
  that are already in the true vacuum).
\end{enumerate}
 
\item evolve the resulting field configuration to the next time step,
  following the field equations (\ref{rho_damp}),
  (\ref{ab_higgs_theta}) and (\ref{ab_higgs_a}) (discretised in the
  gauge-invariant way described in, for example \cite{mmr}), using a
  Runge-Kutta fourth-order algorithm.
\end{enumerate}    

In order to minimise boundary effects we nucleated bubbles only within
a box which was a few bubble radii smaller than our simulation volume,
as was done by Srivastava \cite{srivastava}, and also used this
smaller box to determine when the phase transition was complete.

To generate the bubble nucleation events, we first fix their number,
then choose at random the space-time points at which they take place
within the simulation volume. Since we only nucleate those bubbles
that fall within the false vacuum and discard the rest, at later times
into the transition it will become increasingly difficult for the new
bubbles to find themselves in the false vacuum, and consequently fewer
and fewer bubbles will be nucleated. Also, and although in a realistic
situation one would expect the nucleation rate to vary with the amount
of dissipation present in the system, we kept the nucleation rate
constant throughout the simulation series. This is because we are
concerned primarily with investigating the number of vortices produced
per nucleated bubble, $n_d$ -- since computational resources are
finite, we have chosen not to explore the effect of varying the
nucleation rate at this stage.

The only constraint on the nucleation rate is that we wish to ensure
that on average the bubbles reach their terminal velocity before
colliding.  The bubbles collide, on average, when their radius is of
the order of the initial spatial separation of nucleation events
\begin{equation}\label{t_coll}
t_{\rm coll} = t(r = n_d^{1/\left(d+1\right)}),
\end{equation}
where $d$ is the number of spatial dimensions.

We can parametrise the radius at a given time by three quantities: the
initial radius $r_0$, the terminal velocity $v_{\rm ter}$ and the time
taken to reach terminal velocity $t_{\rm ter}$.  These three
parameters are easily obtained from simulations.  Assuming that the
bubble wall accelerates uniformly up to $v_{\rm ter}$ and then moves
with constant velocity, we get
\begin{equation}\label{r(t)}
r(t) = r_0 + v_{\rm ter}\left(t - \frac{t_{\rm ter}}{2}\right).
\end{equation}
Then the condition $t_{\rm coll} > t_{\rm ter}$ can be expressed
\begin{equation} \label{nuc_rate}
n_d > \left[{v_{\rm ter}\, t_{\rm ter}\over 2} + r_0\right]^{d+1},
\end{equation}
and so the value of $n_d$ used was chosen to satisfy this constraint
for all values of $\Gamma$.

Similarly, we have chosen not to investigate the effect of crossing
the Bogomol'nyi limit -- at the critical coupling $\beta =
\lambda/2e^2 = 1$, a pair of vortices remain stationary (see, for
instance \cite{shellard_vilenkin}).  Above this limit they repel one
another, and below they attract.  We have chosen, for reasons
described in section \ref{sec:params}, parameters corresponding to
$\beta <1$ for our simulations.

\section{Choice of Simulation Parameters} \label{sec:params}

\begin{figure} 
\begin{center}
\epsfig{figure=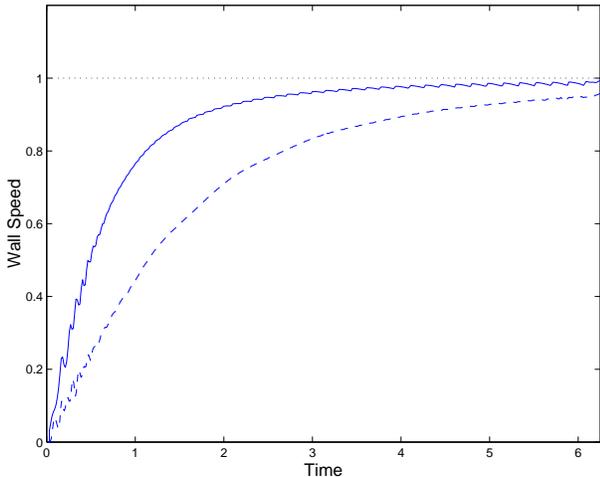, width=8cm} 
\caption{Illustration of effect of asymmetry parameter $\varepsilon$ on
  the acceleration of the bubble wall.  The solid line shows the
  evolution of wall speed with time for $ \varepsilon = 0.4$.  This
  bubble, with a small radius and large difference in energy density
  across its membrane, accelerates rapidly towards the speed of light.
  The dashed line shows the wall speed for $ \varepsilon = 0.1$.  The
  speed of this bubble wall approaches unity much more slowly.}
\label{fig:accel}
\end{center}
\end{figure}

\begin{figure} 
\begin{center}
  \epsfig{figure=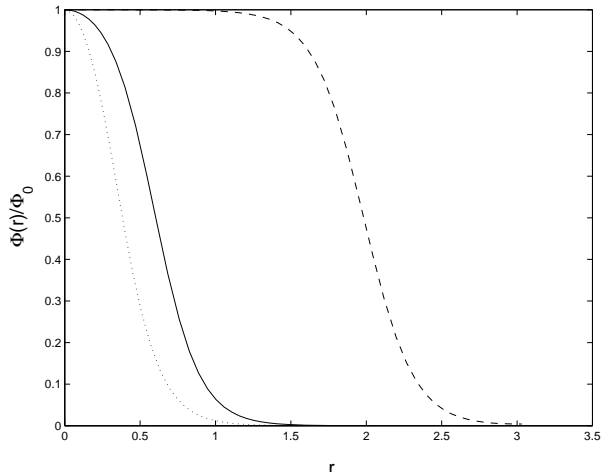, width=8cm} 
\caption{Comparison of wall profiles for different parameters. The
  dashed line shows the profile for $\varepsilon = 0.1$, which has a
  thin wall and large radius. The solid line shows the profile for
  $\varepsilon = 0.4$ and the dotted line that for $\varepsilon =
  0.8$, the value used in this paper.  The radii of the bubbles
  decreases, and the thickness of the wall increases with increasing
  $\varepsilon$, making high values of $\varepsilon$ most suitable for
  these simulations. }
\label{fig:profile}
\end{center}
\end{figure}

The intention of this project is to investigate the effect of bubble
terminal velocity on the defect formation process. It is therefore
necessary to ensure that the bubbles, once nucleated, reach their
terminal velocity $v_{\rm ter}$ as rapidly as possible in order to
decrease the probability of bubbles colliding at speeds lower than
$v_{\rm ter}$ and contaminating our results.  This is achieved by
choosing a relatively high value of the asymmetry parameter,
$\varepsilon=0.8$ -- Figure \ref{fig:accel} shows the effect of
$\varepsilon$ on the acceleration of a bubble. This choice also has
two advantages: the ratio of the radius of a nucleated critical bubble
to its thickness is relatively small (compared to lower values of
$\varepsilon$), which aids numerical accuracy (by decreasing the
spatial resolution required to produce stable results) and it is the
value chosen by Ferrera for his simulations, thus enabling a direct
comparison of the results in global and local-symmetry phase
transitions. Figure \ref{fig:profile} shows the difference in the
bubble wall profile for small and large values of $\varepsilon$.

For the gauge coupling constant, we took $e=0.25$ throughout.  At each
value of $\Gamma$ a number of phase transitions at $e=0$ (i.e a global
symmetry model) were simulated to ensure consistency with the results
of Ferrera (Figure \ref{fig:ferrera}).  Since we primarily wish to
investigate the effect of the bubble wall speed on the number of
defects formed, we have chosen to perform all our calculations with
the conductivity $\sigma=0$.

\section{Ending the Phase Transition}

Vortices have dynamics, and as such it is important that we specify
{\em when} we intend to count them.  If the simulation were allowed to
run indefinitely, all vortices formed would eventually annihilate,
leaving none -- magnetic flux is conserved, and since we started out
with zero total flux, at any given time there must be an equal number
of vortices and antivortices.  Following \cite{ferrera98}, we have
decided to terminate the simulation when 95\% of the simulation volume
had been converted to true vacuum.  As a check, the number of vortices
and antivortices present was calculated at regular intervals and
stored to ensure that a) the number of defects was an increasing
function of time, and b) the number of defects had remained constant
for some time before the phase transition was stopped.  Any
uncertainty introduced by this choice of 95\% completion would have
been compensated by the large number of simulations performed for each
value of $\Gamma$.  Indeed a glance at the cumulative vortex count
shows that the number of defects present in the simulation volume does
not change significantly if the truncation time is increased or
decreased by 10\%.

At the end of the phase transition, the number of vortices and
antivortices produced were counted, using an algorithm which went
across the grid checking at each point for gauge-invariant phase
windings \cite{arttu} of $2\pi n$ around the smallest square going
clockwise from that point with all four corners in true vacuum.  That
the number of vortices and antivortices must be equal is another check
on the accuracy of the code.

\section{Details of Computation}

The first-order phase transition was simulated for 7 values of
$\Gamma$.  For $\Gamma = 0, 2, 3, 5$ and $7.5$, 30 simulations were
performed at each value, whereas for $\Gamma = 10$, only 5 were
performed, due to the increasing amount of time taken to complete the
phase transition.  At each value of $\Gamma$, the simulations were run
on a smaller and smaller lattice spacing with identical initial
conditions, until no divergence in the results could be detected.

A lattice spacing of $0.02$ was determined to be sufficient for all
values of $\Gamma$ and so was used throughout.  For reasons of
simulation time, rather than memory, the spatial lattice size was then
set at $3000^2$ grid points, giving a total physical area of $60^2$.
This box size corresponds to around 20 bubble radii, for the case
where $\varepsilon = 0.8$. The time step $\Delta t$ was set so that it
satisfied the Courant condition (see, for instance p.829 of
\cite{numrec} for further details) for numerical stability -- namely

\begin{equation} \label{courant}
{\Delta t \over \Delta x} \leq {1 \over \sqrt{d}},
\end{equation}
where $d$ is the number of spatial dimensions.  Altogether, we
estimate that this project required around 1300 hours of combined CPU
time on the supercomputer.

\begin{figure*}
\begin{center}
  \epsfig{file=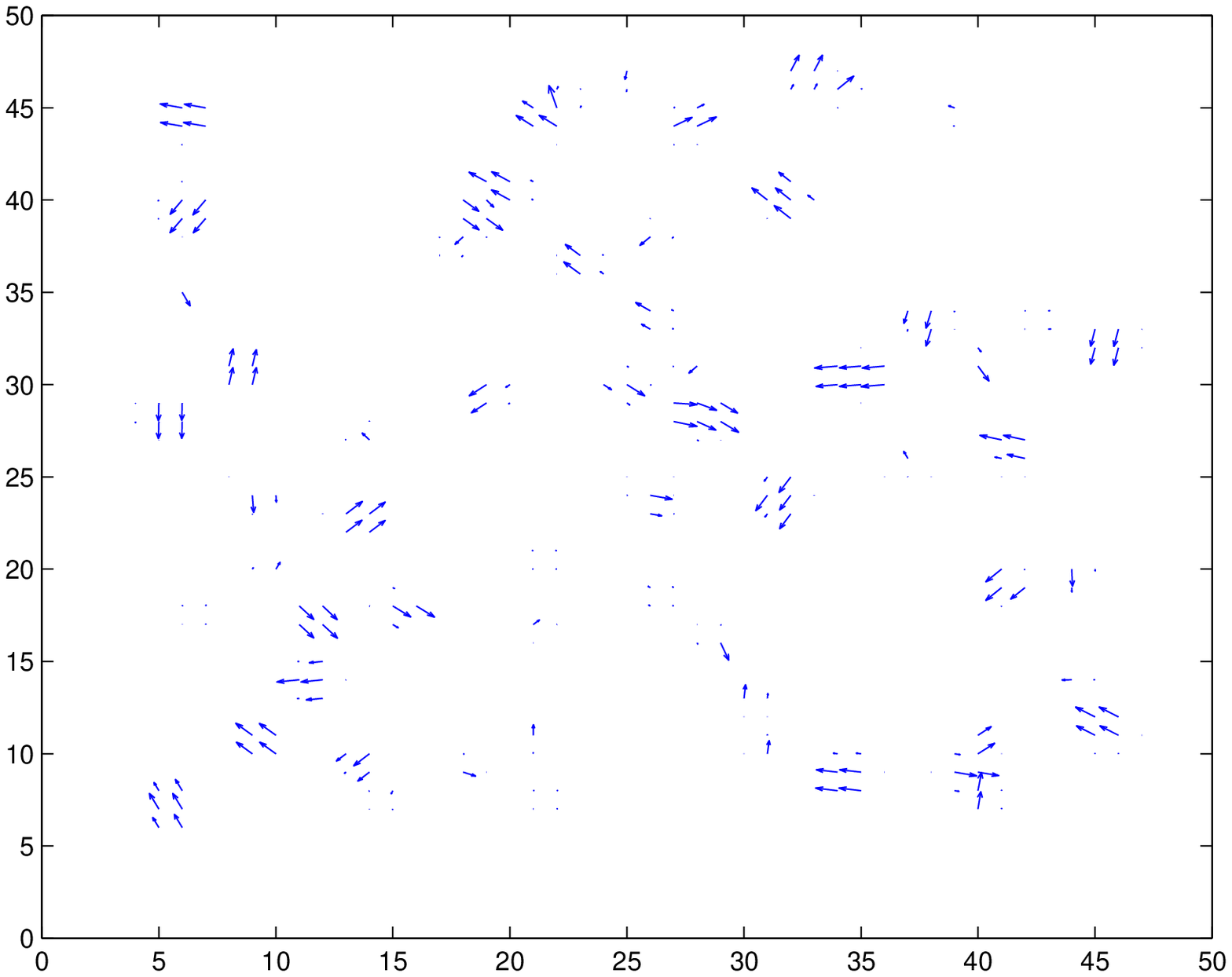, width=4.3cm}
  \epsfig{file=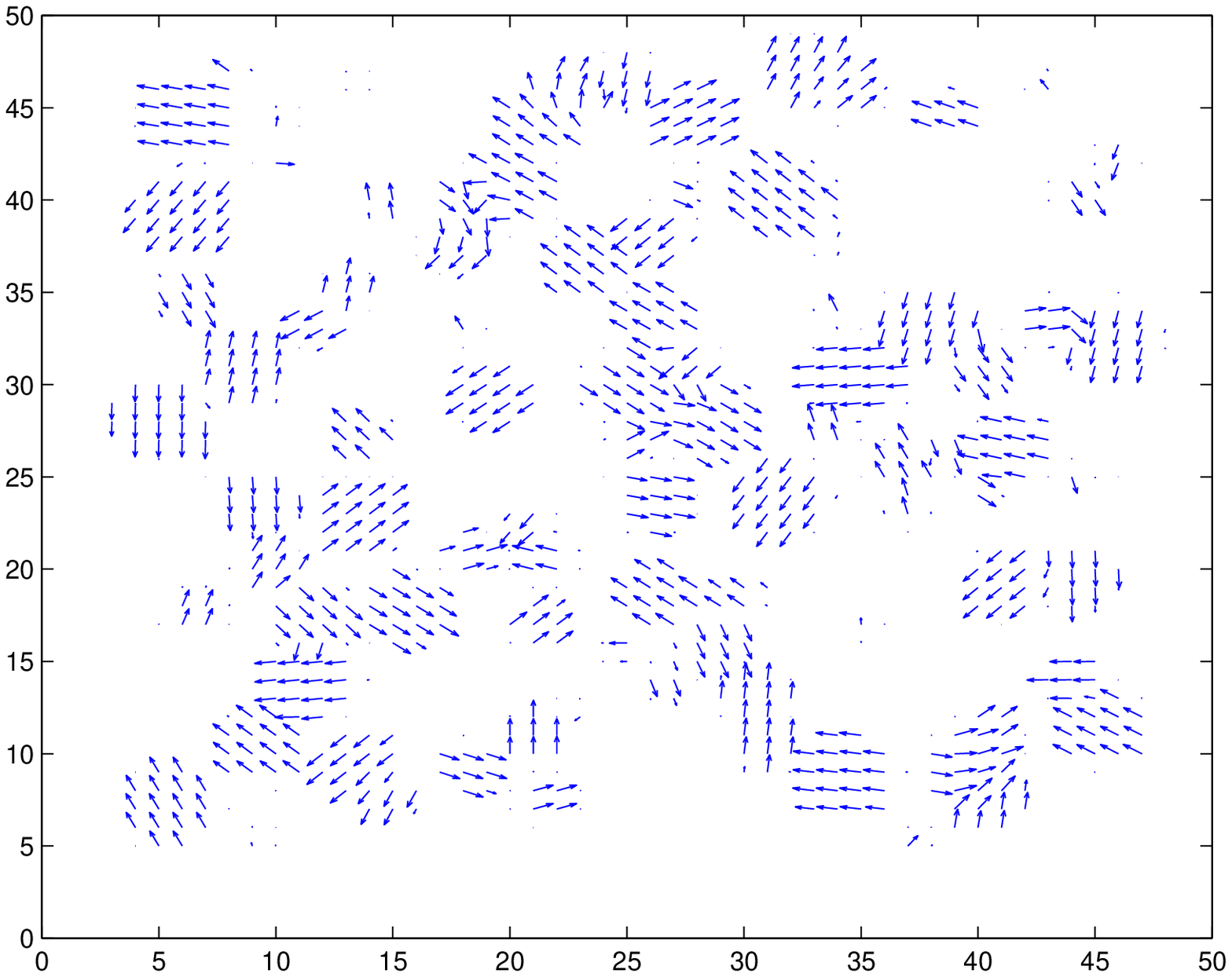,width=4.3cm}
  \epsfig{file=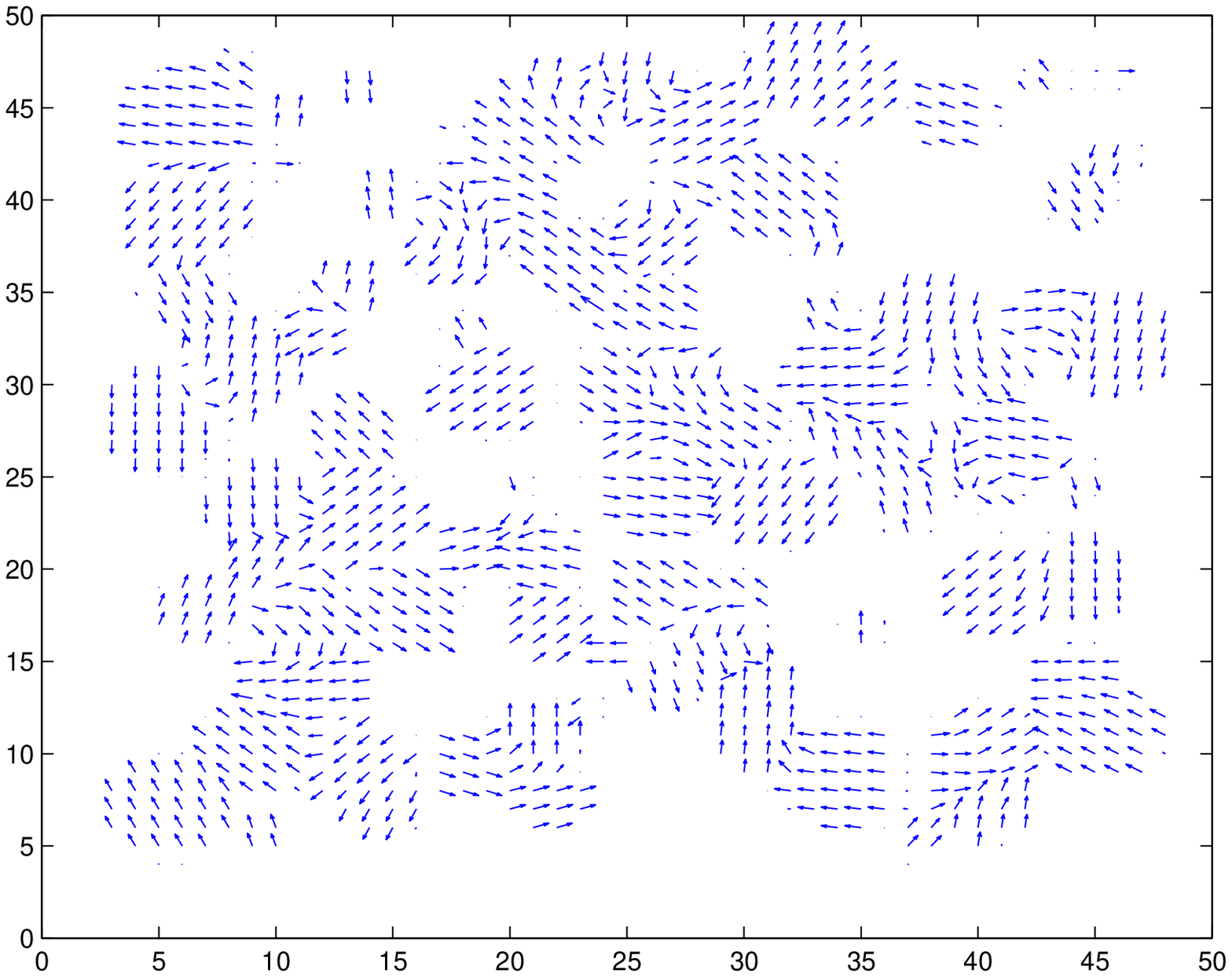, width=4.3cm}
  \epsfig{file=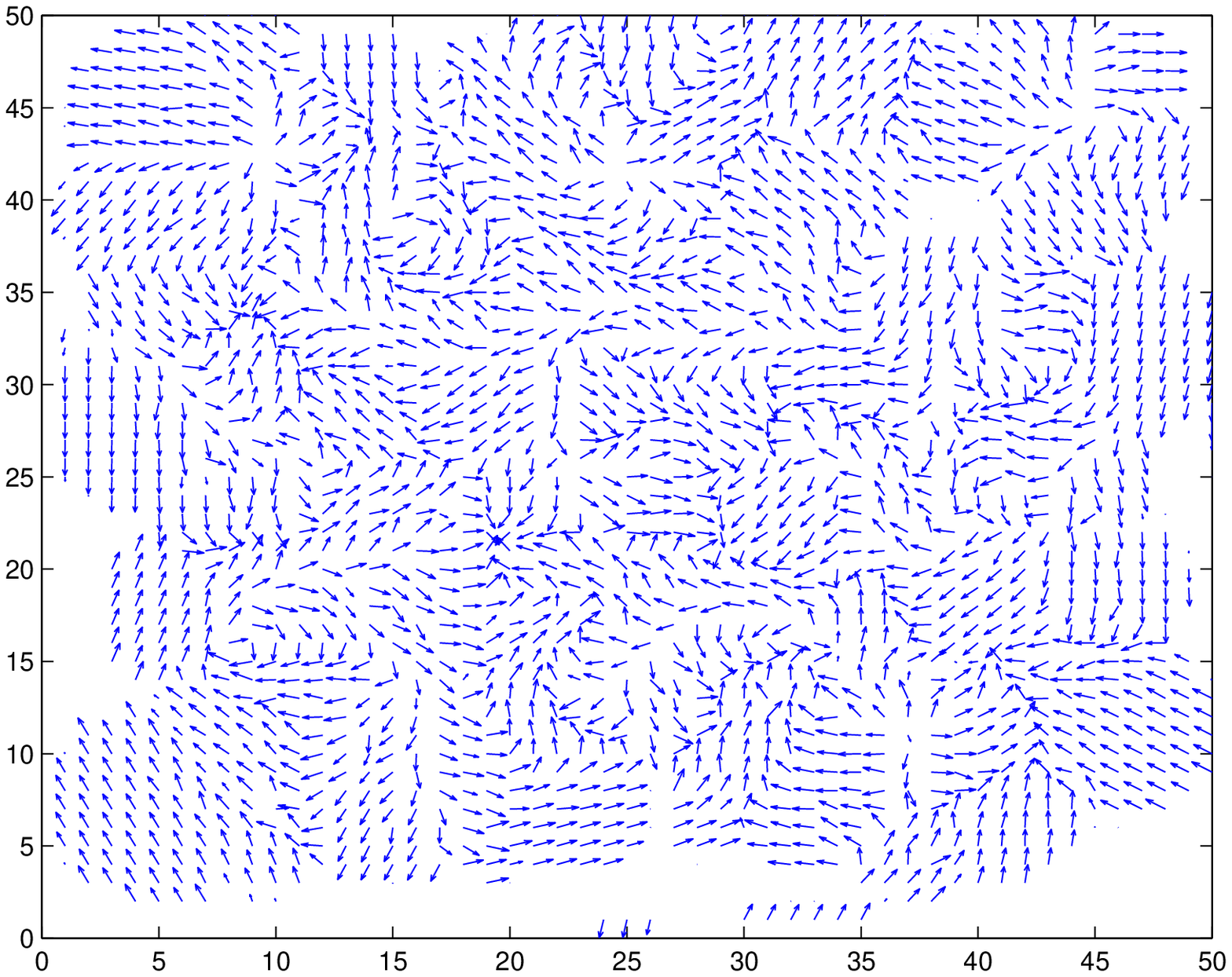,width=4.3cm}
\vskip 0.2 in
\caption{Snapshots of a phase transition with $\Gamma = 0$ at
  different times.  From top left, they are at $t= 2.2, 3.67, 5.13,
  7.33$, and the phase transition was declared complete by the program
  at $t=7.80$.  The length of the arrows is proportional to the value
  of $|\Phi| = \rho$, and their direction gives the phase $\theta$ of
  the Higgs field.  In particular, the false vacuum (which is $\Phi =
  0$) corresponds to an absence of lines.  The first image is very
  early in the phase transition, when only very few bubbles have been
  nucleated.  These bubbles expand, alongside new bubbles which
  continue to be nucleated. In the second and third of these images,
  the existence at a given instant of bubbles of differing sizes is
  evident.  The final image shows the phase transition almost
  complete.}\label{fig:pretty_pictures}
\end{center}
\end{figure*}

\section{Simulation Results}

\begin{figure}
\begin{center}
  \epsfig{file=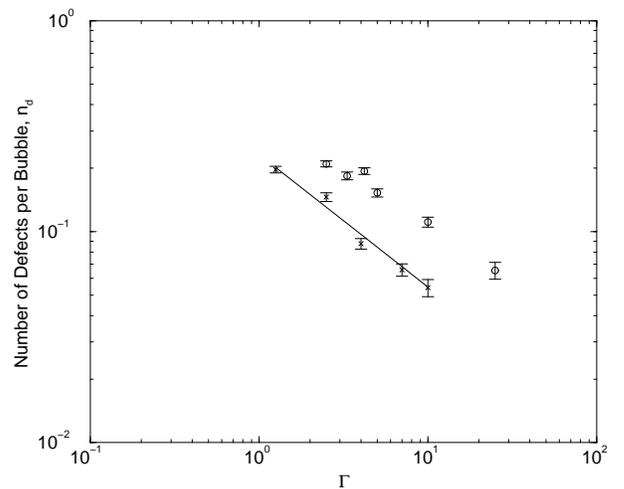 ,width=8cm}
\caption{Log-log plot of number of defects formed per bubble nucleated $n_d$
  vs. the friction coefficient $\Gamma$.  The lower set of points
  (data points marked with a cross) are the results of this
  investigation, the upper set (marked with a circle) are those
  obtained by Ferrera \protect\cite{ferrera98}.  The two hypotheses of
  \protect\cite{lilley_bubble} are demonstrably supported -- namely i)
  that $n_d$ should decrease with increasing $\Gamma$, and ii) that for a
  given value of $\Gamma$, $n_d$ should be lower for $e=0.25$ (the
  gauge symmetry) than for $e=0$ the global symmetry. The solid line
  corresponds to the least-squares, best-fit for $n_d \propto
  \Gamma^\alpha$, with $\alpha = -0.65\pm 0.04$.}\label{fig:sim_graph}
\end{center}
\end{figure}

The main result of our simulations can be seen in Figure
\ref{fig:sim_graph}.  Our conclusions are twofold and can be stated
thus -- the number of topological defects formed per bubble nucleated
{\em did} decrease significantly as the terminal velocity of the wall
decreased; the number of topological defects formed per bubble
nucleated {\em was} significantly lower for gauge theories than for
global theories.  Both of these statements support the hypotheses made
in \cite{lilley_bubble}.

We have shown that the number of defects formed per bubble nucleated,
$n_d$, decreases exponentially with bubble wall velocity (since
$v_{\rm ter} \sim 1 / \Gamma$). The cosmologically-relevant quantity
would be, of course, the number of defects per unit volume, $n_v$.
These are related by
\begin{equation} \label{n_d}
n_v = {n_d \over {\mbox{number of bubbles per unit volume}}} \sim {n_d
  \over r^{-3}}, 
\end{equation}
where $r$ is the average bubble radius on collision.  Since the only
parameters in this system are the damping coefficient $\Gamma$ and the
bubble nucleation rate per unit time per unit volume $N$, we have 
\begin{equation}\label{radius}
r \sim [N \Gamma]^{-{1\over d+1}},
\end{equation}
where $d$ is the number of spatial dimensions.  If, as we have
discovered, $n_d \sim \Gamma^{-\alpha}$, we have
\begin{equation}\label{num_vol}
n_v \sim N^{3\over d+1} \ \Gamma^{\left({3 \over d+1} - \alpha\right)}.
\end{equation}
If we assume a fixed nucleation rate, slow-moving bubbles have two
effects with respect to the formation of topological defects.  The
first is that more bubbles must be nucleated to complete the phase
transition, and the second is that phase equilibration suppresses the
formation of defects in some cases.  Our results indicate, since we
have found $\alpha < 1$ for $d=2$, that the increase in the number of
bubbles nucleated outstrips the decrease in the defect-formation rate
due to phase equilibration, and that slow-moving bubbles lead to {\em
  more} defects in any given volume.  

Here we must point out that this statement is only valid for the range
of wall velocities we have considered $(0.05 \lsim v_{\rm ter}/c \leq
1)$.  In this range, we have measured the exponent $\alpha_{\rm local}
= 0.65 \pm 0.04$, whereas in the global case \cite{ferrera98}
$\alpha_{\rm global} = 0.53 \pm 0.05$ was found for the corresponding
region.  For lower terminal velocities, however, another regime was
discovered, with $\alpha_{\rm global} = 0.8 \pm 0.3$.  The crossover
between the two regimes corresponded to a bubble velocity low enough
that the time taken for the bubble walls to merge on collision was
longer than the time light took to cross the bubble.  We have not been
able to probe this velocity region, and cannot rule out the
possibility that for very slowly-moving bubble walls, defect
suppression due to phase equilibration will outstrip the effect of
more bubbles being nucleated, and so lead to fewer defects per unit
volume.

\subsection{Results for $\Gamma=0$}

Not shown on Figure \ref{fig:sim_graph} (which is drawn with log-log
axes) is the result for $\Gamma = 0$ -- for this we obtained $n_d =
0.240 \pm 0.02$. This agrees well with the prediction of the Kibble
mechanism, that on purely geometrical grounds the number of defects
formed per bubble should be $0.25$.  As stated in the Introduction,
Melfo and Perivolaropoulos \cite{MP} found that phase equilibration
due to non-simultaneous three-bubble collisions did {\em not}
significantly alter the defect formation rate for global symmetries
($e=0$), with bubbles moving at the speed of light ($\Gamma=0$).

We can now confirm that in our gauge-theory model, with bubbles moving
at the speed of light, we find no significant deviation from the
Kibble mechanism prediction.  This result, we feel, is significant
enough to be stated in its own section.

\section{Summary}

In this paper we have given an account of our attempts to simulate a
first-order phase transition incorporating one of the effects of an
early-universe environment -- a retarded bubble wall.  We have found,
as we predicted, that the number of topological defects formed per
bubble nucleated, $n_d$, decreases with the terminal velocity of the
bubble wall.  Moreover, the number of topological defects produced is
significantly lower, for every value of $\Gamma$, in the gauge theory
than in the global theory.

Slow-moving bubbles allow, in general, more time between collisions
and so facilitate the suppression of defect formation due to phase
equilibration.  Another effect of slow-moving bubbles, however, is the
need for more bubbles to be nucleated in order for the phase
transition to complete.  Our studies show that the surfeit of bubbles
dominates the suppression of defects, and that slow-moving bubbles
$(0.05 \lsim v_{\rm ter}/c \leq 1)$ actually lead to more defects in a given
volume.

This investigation is far from complete.  We would have liked to have
been able to probe the region $v_{\rm ter} < 0.01$, as this may be
more realistic for cosmological scenarios.  We would also like to have
investigated the effect of plasma conductivity on the defect formation
rates, both separately from and in tandem with the probe on the effect
of wall terminal velocity.  Unfortunately, time and computational
constraints did not allow this up until now.

Of course, any such cosmological phase transition would take place not
in a static background, but in an expanding universe.  The expansion
of the Universe has not been taken into account here, but could be
extremely important for any high-energy (e.g. GUT-scale) phase
transition.  This, along with the low-velocity and high-conductivity
regimes, is something we hope to probe in the future.


\section{Acknowledgements}
We would like to thank A.C. Davis, S. Gratton, W. Perkins, P. Saffin,
P. Shellard and T.  Wiseman for helpful comments and conversations.
Computer facilities were provided by the UK National Cosmology
Supercomputing Centre in cooperation with Silicon Graphics/Cray
Research, supported by HEFCE and PPARC.  This work was supported in
part by PPARC and an ESF network grant.  We would also like to thank
CSIC in Madrid, where this project was begun, for hospitality.
Support for M.L.  was provided by a PPARC studentship and Fitzwilliam
College, Cambridge.

\end{document}